# Information, Community, and Action: How Nonprofit Organizations Use Social Media[*]

**Kristen Lovejoy**
Department of Communication, University at Buffalo, SUNY

**Gregory D. Saxton**
Department of Communication, University at Buffalo, SUNY

*Abstract*

*The rapid diffusion of "micro-blogging" services such as Twitter is ushering in a new era of possibilities for organizations to communicate with and engage their core stakeholders and the general public. To enhance understanding of the communicative functions micro-blogging serves for organizations, this study examines the Twitter utilization practices of the 100 largest nonprofit organizations in the United States. The analysis reveals there are three key functions of micro-blogging updates —"information," "community," and "action." Though the informational use of micro-blogging is extensive, nonprofit organizations are better at using Twitter to strategically engage their stakeholders via dialogic and community-building practices than they have been with traditional websites. The adoption of social media appears to have engendered new paradigms of public engagement.*

---

[*] Authors are listed in alphabetical order. The authors would like to thank Tom Feeley, Richard Waters, Seungahn Nah, I-hsuan Chiu, Yuchao Huang, and Kenton Anderson for helpful comments and suggestions.

**Information, Community, and Action: How Nonprofit Organizations Use Social Media**

Prior studies (e.g., Kent, Taylor, & White, 2003; Saxton, Guo, & Brown, 2007) have shown that nonprofit organizations have not been able to use websites as strategic, interactive stakeholder engagement tools. Perhaps this was due to not having the know-how or the staff to create more interactive sites with feedback options and discussion boards. The advent of social networking sites like Facebook and Twitter have taken away this excuse. These sites are free and have built-in interactivity. Any organization big or small can create a site and start building a network of friends and followers with whom they are in almost real-time contact.

These newer social media applications present communication opportunities that differ dramatically from organizationally supported websites. The question, is, how are organizations using micro-blogging applications? We aim here to help answer this question and understand the various functions for which organizational micro-blogging is employed via an examination of the Twitter utilization practices of the 100 largest charitable organizations in the United States. First, based on an analysis of the tweets sent by the organizations in our sample, we develop an original typology of organizational micro-blogging functions. Second, we use this information to examine the relative frequency with which organizations rely on the main types of tweets, and to then characterize the organizations themselves based on their tweet function utilization patterns.

Our plan for the paper is as follows. After reviewing the literature and presenting our theoretical framework, we describe our sample and coding procedures and then turn to an examination of our findings. We end with a discussion of the core implications of our study for organizations, for civil society, and for theory. In the end, scholars will have to continue to develop new conceptual tools to understand stakeholder relations in an age dominated by social media.

**New Media as an Organizational Communication and Stakeholder Engagement Vehicle**

Though new media have certainly not replaced more traditional communication technologies in the workplace, the digital continues to crowd out the analogue: Of 17 workplace communication technologies considered "mandatory" by 50% or more of respondents in a recent survey, ten were computer-mediated, including email, the Internet, collaboration software, wikis, and Web teleconferencing (D'Urso & Pierce, 2009). As these and newer media spread, an increasing proportion of organizational models, operations, and transactions are purely electronic (e.g., Kang & Norton, 2004). The spread of new media has also significantly increased nonprofits' ability to communicate with clients as well as regulators, volunteers, the media, and the general public (e.g., Waters, 2007). Through strategically targeted content, firms can mobilize stakeholders, build meaningful relationships, and ultimately foster increased accountability and public trust (Saxton & Guo, 2011). Online nonprofit/stakeholder interactions have effectively become more and more ubiquitous, multifaceted, and critical to organizational performance.

A burgeoning line of inquiry has thus set out to investigate the various ways organizations are communicating with and engaging the public via new media. The first generation of studies shed light on the tools of Web 1.0, primarily websites. Scholars have examined online dialogic relationship-building in diverse settings, including community colleges (McAllister-Spooner & Taylor, 2007), for-profit corporations (Park & Reber, 2008), and not-for-profits (Kent et al., 2003), and have examined the use of websites to boost stakeholder responsiveness (Saxton, Guo, & Brown, 2007), generate social capital (Nah, 2009), and develop strategic and interactive stakeholder communications (Hackler & Saxton, 2007; Waters, 2007).

These website studies collectively point to two broad organizational functions for new media: information and dialogue. These studies further suggest the website has become both the "public face" of the organization and the vehicle through which intense and meaningful public interactions can take place. What the first generation of studies showed, however, was that despite the promise Web 1.0 technologies sparked to engender a more active and "dialogic" civil society, nonprofits have been unable to use the website as a strategic, interactive organizational communication and stakeholder engagement tool (e.g., Kent et al., 2003; Saxton et al., 2007).

**Social media and organizational communication**

The development of newer forms of communication technologies, such as blogs, instant messaging, and Internet Relay Chat, heralded a qualitative change in communicative potential beyond that possible through email or websites. Studies of these newer digital technologies, which could be considered "precursors" to social media, found they engendered substantially greater capabilities for collaboration, interactivity, and "polychronic" communication (Cameron & Webster, 2005; Macias, Hilyard, & Freimuth, 2009; Quan-Haase, Cothrel, & Wellman, 2005).

The advent of social media has opened up even greater possibilities for interpersonal and organizational communication. At the interpersonal level, scholars have examined the role of Facebook in building social capital (Ellison, Steinfield, & Lampe, 2007). Twitter has also been the focus of a growing body of interpersonal research (e.g., Java, Song, Finin, & Tseng, 2007; Naaman, Boase, & Lai, 2010). While they provide evidence of a connection between individual users' tweet profiles and community formation (Java et al., 2007), these studies do not so much demonstrate the positive social effects of social media so much as the ability of Twitter to serve as a vehicle for narcissism, opinion-making, and information-sharing. By contrast, studies by Hughes and Palen (2009) and Smith (2010) have shown how Twitter can serve as a valuable communication and information-sharing resource during emergency-relief efforts.

Organizational-level research on social media has not grown as rapidly. What little research that does exist on nonprofit Facebook utilization (Bortree & Seltzer, 2009; Greenberg &

MacAulay, 2009; Waters, Burnett, Lamm, & Lucas, 2009) regards the heavy reliance on basic informational uses as a lost opportunity for furthering interactivity and dialogue with supporters.

The organizational-level research on Twitter utilization is similarly sparse. Briones, Kuch, Liu, & Jin's (2011) interviews with 40 Red Cross employees showed support for the relationship-building potential of Facebook and Twitter. Rybalko and Seltzer (2010) found evidence of "dialogue" in *Fortune 500* companies' Twitter accounts. And Jansen, Zhang, Sobel, and Chowdury (2009) showed companies have a different set of priorities than individuals. Organizations were less likely to send "me now"-type tweets, more likely to send informational messages, and more likely to try to engage their followers with messages that spurred dialogue.

**Classifying Social Media Messages**

This study concentrates on a central feature of organizations' social media utilization—the actual messages sent. This is distinct from the other primary feature of social media applications—the inter-personal and inter-organizational network structures developed via friending and following behavior. To date, no published papers exist that attempt to code Facebook messages. As for Twitter, several attempts at tweet categorization have been made (Java et al., 2007; Naaman et al., 2010), but none have looked at organizational applications. These studies showed that, at the individual level, the "daily chatter," or "me now," type of tweets were the most prevalent.

**Summary: New Media and Organizational Communication Functions**

Overall, social media appears to have created opportunities for interpersonal engagement, interactivity, and dialogue that are qualitatively different from those offered by traditional websites. It would thus be reasonable to infer that social media would similarly carry considerable potential as an organizational communication and stakeholder relations tool. However, prior research only hints at the scope and nature of stakeholder-related organizational communications made possible by the diffusion of social media. Though the more extensive interpersonal-level social media research likely entails hypotheses that are testable at the organizational level, only the handful of preliminary studies just noted explicitly discuss how organizations use social media to communicate with their stakeholders and the general public. There is a pressing need for both relevant data and analytical frameworks that can help us understand how organizations are using microblogging to engage the public.

A key insight found in the individual-level schemes noted above, in the existing organizational-level studies on website utilization (e.g., Kang & Norton, 2004; Kent et al., 2003; Waters, 2007), and in the broad findings of the existing organizational-level social media studies (Jansen et al., 2009; Waters et al., 2009) is that organizations seem to employ new media for at least two primary purposes—information-sharing and dialogic relationship-building.

The problem is that, broadly speaking, we do not yet have a good sense of how organizations are using social media. We do not know whether and how the information and dialogic functions manifest themselves on organizations' social media sites. We do not know what new forms of communication might be being utilized by organizations, nonprofit or otherwise. And we do not know which of the principal forms of organizational communication are most prevalent and central to the organizational mission. We aim to help shed light on these issues in proposing the two following research questions:

> *Research Question 1:* How are organizations using micro-blogging applications? More specifically, for what functions is organizational micro-blogging being employed?

> *Research Question 2:* How do organizations vary in their reliance on the primary micro-blogging functions?

## Method

### Sample

To address these questions, we employ both tweet-level and organizational-level analyses of 100 nonprofit organizations' use of Twitter. Twitter was launched in October 2006 and has since become the largest micro-blogging site on the Internet, making it an excellent way to reach a large number of stakeholders with brief, easily digestible bits of information. Only messages of 140 characters or less can be sent at a time; these messages, called "tweets," are broadcast in real-time and become part of the public stream unless a user sets her tweets to a private setting.

Given the novelty of social media applications as organizational communication tools, for our sampling frame we utilized a sample of large charitable organizations, which despite the low financial and technological barriers to adoption are still more likely than smaller organizations to have a significant presence on Twitter. Specifically, as in prior studies (e.g., Kang & Norton, 2004), we examined organizations from the most recent version (2008) of the "*Nonprofit Times* 100" list available at the start of our study period. Published annually in the *NonProfit Times*, the list contains the 100 largest non-educational US nonprofit organizations in terms of revenue. To make the list, at least ten percent of revenues must come from donations.

**Data-gathering and Sample Characteristics**

To gather data on these organizations, our first step was to determine which organizations had Twitter accounts. To find active accounts, we undertook a multi-pronged search strategy during the first week of November 2009. First, we searched the organizations' websites: most of the organizations (55) had a prominent link to their Twitter account on the home page, while another 3 organizations had a link to a Twitter account on a sub-page of the website. To find the accounts of the remaining organizations, we conducted a search on both Google (e.g., "Twitter Boy Scouts of America") and the Twitter search engine. This yielded an additional 15 organizations.

In sum, 73 of the 100 organizations were found to have Twitter accounts. The organizations represented a cross-section of the charitable sector. In terms of the National Taxonomy of Exempt Organizations (NTEE) classification system codes, 27% of the organizations were operating in the field of International and Foreign Affairs, 15% in Arts, Culture, and Humanities, 24% in Health, and 8% in Youth Development, while the remainder operated in a variety of other fields, including the environment, public safety, human services, recreation and sports, housing and shelter, and mental health and crisis intervention, among others. A Web Appendix (available at http://www.gregorysaxton.net) contains a complete list of the organizations with associated NTEE codes.

Subsequently, we began to gather Twitter utilization data on these organizations for the month-long period from November $8^{th}$ to December $7^{th}$, 2009. All organizational tweets published during this period were downloaded into an *SQLite* relational database via the Twitter application programming interface (API), using Python code written specifically for this research (available upon request). The final database contained 4,655 tweets, which were doubled-checked against the Twitter stream for 10 of the organizations and found to be complete in all cases. On average, an organization sent out 66 tweets over the 30-day period (s.d. = 65.74), with the frequency ranging from 0 (in the case of one organization) to 289.

**Code development**

Our main task in this paper is to analyze the content of organizations' tweets and determine what communicative function they serve. For this analysis, we develop an original micro-blogging function categorization scheme. As noted above, several studies have classified messages sent by individuals on Twitter; ours is the first to classify social media messages by *organizations*.

The classification scheme we developed was informed by previous individual-level social media coding schemes, prior blog classification studies (e.g., Macias et al. 2009), and the new media and stakeholder engagement literatures. Building on prior research, we thus began deductively with the assumption that we would likely find informational and dialogic forms of communication in our sample of organizations. However, ultimately, the codes were developed via an inductive process based on a review of tweets from a sample of nonprofit organizations in the month prior to our data-gathering period (October 2009).

**Coding procedures and inter-coder reliability**

Our database contained the 4,655 tweets sent between November 8 and December 7, 2009 by the 73 organizations with Twitter accounts. Given the large number of tweets, the decision was made to code a subset—the 2,437 tweets sent over the first two weeks of the study. Both authors began by coding the first 100 tweets using the 12-category scheme shown in Table 1; each tweet was assigned a single code from this scheme. In cases where a tweet appeared to serve dual purposes, codes were assigned according to what was considered the tweet's primary purpose. Discrepancies between codings were discussed and coding rules refined until 100% agreement was reached. Using the refined rules, another 100 tweets were coded with 94.0% inter-coder agreement and a Cohen's kappa score of .91, indicating a high level of inter-coder reliability.

# Results

In line with our two primary research questions, our analysis of stakeholder communication on Twitter has two components. First, we examine the content of the aggregate set of tweets via our original typology of organizational micro-blogging functions. We then analyze and classify the organizations based on their relative utilization of the various micro-blogging functions.

**Functions of organizational micro-blogging**

As shown in Table 1 and discussed in detail below, twelve types of tweets emerged from the coding process. Based on our inductive analysis of the data, we then grouped these categories into three major functions: Information, Community, and Action. By way of initial introduction, we might think of informing as the basic function of Twitter; this involves spreading information about the organization, its activities, or anything of potential interest to followers. The second function, "community," taps into how organizations can foster relationships, create networks, and build communities on Twitter through tweets that promote interactivity and dialogue. The heart of this function are "dialogic" messages and those that attempt to build a community of followers via "bonding" messages, such as "thank you" and acknowledgement tweets. The third function, which we call "action," has as a central purpose the aim of getting followers to "do something" for the organization, whether it is to donate, buy a product, attend an event, join a movement, or launch a protest. Promotion and mobilization are at the heart of this function.

We now turn to our analysis of the twelve specific tweet categories we developed, organized according to primary function.

[Insert Table 1 here]

**Information**

The information function contains a single category, which covers tweets containing information about the organization's activities, highlights from events, or any other news, facts, reports or information relevant to an organization's stakeholders. In line with previous organizational website research (e.g., Saxton et al., 2007; Waters, 2007), it involves a one-way interaction, the exchange of information from the organization to the public. Interestingly, a large proportion of tweets in this category moved beyond simple "tidbits" of information amenable to a 140-character limitation. As seen in the following example (with organization name shown in italics), many of these tweets included links to other sites where additional information could be found:

> *NYPL:* Phillip Hoose is on stage, introducing Claudette Colvin! Awesome! #nationalbook http://yfrog.com/31nc8j

The main difference between this category and the others is that the tweet's primary purpose is solely to inform; there is no explicit secondary agenda, that is, the tweet does not chiefly serve to promote an event, mobilize supporters to take some type of action, foster dialogue, or build a community. What binds these tweets in common is the simple one-way information exchange.

As shown in Table 1, most messages (59%) were classified as informational. This is consistent with prior research (Greenberg & MacAulay, 2009; Jansen et al., 2009). While much of the literature implicitly maligns the informational function (e.g., the pejorative label "brochureware"), the effects of democratized information-sharing processes can be far-reaching. Information is a powerful tool during crises (Macias et al., 2009), and at other times, when an organization sends information about its activities or its history, vision, or objectives, or detailed information on its finances, performance, governance policies, or ethical standards, it can connect a broad array of stakeholders to its mission and help to boost accountability and public trust. Informational tweets can also help connect an organizations' constituents to relevant resources in the community. At the same time, informational tweets serve as an essential base upon which more complex functions (e.g., dialogue and mobilization) can be built.

**Community**

The current literature on social media use by nonprofits shows the gap between sending out information and creating "dialogue" (e.g., Waters et al., 2009). Although Twitter's main purpose is micro-blogging, it is also a social networking tool. Organizations can thus use Twitter to interact, share, and converse with stakeholders in a way that ultimately facilitates the creation of an online *community* with its followers. We label tweets that fill this function "community." There are effectively two aspects to this function: dialogue and community-building. First, there are tweets that spark direct *interactive* conversations between organizations and their publics; this is similar to the notion of "dialogue" in the organizational website literature (e.g., Kent et al., 2003). Second, there are those tweets whose primary purpose is to say something that strengthens ties to the online community without involving an expectation of interactive conversation. This element relates to the social capital and network-building functions that Nah (2009) suggests is possible in organizational websites.

We found four categories of tweets that fulfill the community function. Two of the categories, *giving recognition and thanks* and *acknowledgment of current and local events,* are primarily related to the "community-building" element, while "responding to public reply messages" and "response solicitation" are more directly associated with the "dialogue" aspect. What binds all four categories is the goal of building and engaging with a productive and healthy online community comprising the organization and its supporters. Altogether, 26% of the tweets in our sample served the dialogic and community-building function.

*Giving recognition and thanks.* It is one of the basic tenets of nonprofit management that acknowledging and thanking donors and other supporters is essential. Messages giving such thanks accounted for 13.2% of all tweets. As typified by the following example (with "@UserID" shown to protect privacy), it includes giving thanks and recognition to volunteers and sponsors, to followers who had Retweeted the organization's messages or mentioned the organization in their tweets, and to followers who had participated in online contests:

> *ChildrensLA:* We love this tweet!  RT @UserID: This is why we are working with CHLA! Great facilities & docs helping children. http://bit.ly/3I6YxS

*Acknowledgement of current & local events.* This category covers the acknowledgement of noteworthy events, including holiday greetings and support of community events or sports teams. This is a useful way to show the organization is a good neighbor and part of the community, in the same way that, offline, it would be odd to have a conversation with a neighbor on New Year's Eve without any mention of the holiday. Such tweets are also an easy way to spark conversation. Note that, in some cases, tweets containing "acknowledgements" were

HOW NONPROFIT ORGANIZATIONS USE SOCIAL MEDIA  13primarily promotional in intent, in which case they were coded under "action." For example, "Patience and Fortitude Salute the Troops on Veterans Day! The Library Lions love a parade! http://bit.ly/2uvxm5" would be counted as an acknowledgement of Veterans Day, but "Tomorrow, Nov. 11, is Veterans Day. Take a look at some of the services we provide …. http://bit.ly/46fd1E" would not. In any case, this category was not common (9 total tweets).

*Responses to public reply messages.* Although Twitter has a way to send direct, private messages, called "direct messages," the norm is to post a tweet with the "@" symbol before the name of the Twitter user the message is intended for. For example, a user could send a message to the Red Cross by writing "@Redcross" at the beginning of a tweet. Red Cross will be able to see that a user has posted about them in their sidebar. The organization could then reply in the same manner using the "@" symbol. It is the Twitter equivalent of the Facebook "wall" function. Our preferred term for such tweets is "public reply messages." We found that organizations' responses to public reply messages comprised 8% of all tweets in our dataset. For example:

> *PBS:* @UserID hooray! Hope you can join us for the live chat this afternoon. If not, send me a question here and I can pass it along ^LS

Tweets in this category, along with the following, are the clearest expression of "dialogue" between an organization and its stakeholders.

*Response Solicitation.* Our final category in this function involves tweets that solicit a conversational response from stakeholders. Such tweets are important because they clearly show that the organization is looking to create dialogue. These are not just interesting statements that might spark a conversation, but messages that explicitly seek a response of some sort, including polls, surveys, contests, direct questions to followers, and requests to Retweet:

> *NYPL:* TRIVIA: Who wrote and originally recorded the song "Black Magic Woman"? HInt: http://bit.ly/5n43m2

Such messages comprised 4% of all tweets in our study.

We can conclude that Twitter seems to be a more effective dialogic communication tool than the traditional website. In an extensive review, Waters (2007) found that the most common way nonprofits use their website to promote dialogue is by simply collecting e-mail addresses, noting that they rarely use more interactive features such as discussion forums or live chat. This is in stark contrast to the dialogic uses of Twitter. As we have shown above, Twitter by its nature allows for more opportunities for direct interactivity, two-way exchange of information, network creation, and public, open dialogue.

**Action**

The third and final primary function is "action." The heart of this function are messages that aim to get followers to "do something" for the organization—anything from donating money or buying T-shirts to attending events and engaging in advocacy campaigns. It involves the promotional and mobilizational uses of social media messages where, implicitly at least, Twitter users are seen as a resource that can be mobilized to help the organization fulfill its mission.

This function is perhaps the most tangible, outcomes-oriented manifestation of the benefits rendered possible by a Twitter presence, asking followers to do something concrete to help the organization meet its objectives. As a result, this may be what many organizations ultimately want to achieve. They want to mobilize followers to attend events, make donations, and become activists. They want to move their followers, in effect, from informed individuals to members of a community to activists and donors. It is less about creating dialogue than it is about mobilizing resources and supporters to fulfill financial and strategic goals.

The action function includes seven categories of tweets, which we discuss in turn below. Collectively, they comprise 15.6% of all messages sent.

*Promote an event.* Twitter can be an effective tool for promoting events. In fact, this was by far the most common type of action tweet in our dataset, comprising exactly half of all action-oriented messages and 7.8% of all tweets. These tweets did not just include *information* on the event, which would have been put in the "informational" category, but also included a date, time, or price. The promotion is hence explicit and the primary purpose of the tweet:

> *UJAFederation:* Shabbat Sholom, friends & followers :) Candle lighting this eve is at 4:31pm.  Torah portion is Parshat Toldot. http://fbpage.com/toldot

*Donation appeal.* The second most common action message (3.1% of total) was the donation appeal. The way to get a donation is to ask for one, and savvy organizations are using Twitter to make that ask. Messages in this category either directly asked for a donation or asked followers to support companies that were donating a percentage of their sales. For example:

> *Vol_of_America:* Hunger is a daily reality for many people. Help us provide meals & services for needy people this holiday season http://voa.org/caringgifts

*Selling a Product.* Like asking for a donation, the direct selling of a product is another way for an organization to make money using Twitter. For example:

> *MetOpera:* The new online shop is now open! Browse through for great gifts, CDs, DVDs and more! http://www.metoperashop.org

Somewhat surprisingly, such appeals did not occur often in our sample (12 out of 2,437 tweets).

*Call for volunteers and employees.* Organizations also used Twitter to issue calls for employees or volunteers:

> *ChildrensLA:* Pls RT Mission Critical:  Looking for a great online communications coordinator to help our hospital at http://bit.ly/9wRW4 #jobs

In Waters et al.'s (2009) study of Facebook, 13% of the organizations listed volunteer opportunities. In our sample, 15 organizations, or about 20% of the 73 organizations with Twitter accounts, listed volunteer opportunities at one point during the study period. However, such messages comprised less than 1% of coded tweets.

*Lobbying and advocacy.* Twitter could with little technical difficulty be used to directly ask followers to perform a lobbying- or advocacy-related activity, but only 5 organizations in this sample did so. For example, World Vision USA issued the following call:

> *WorldVisionUSA:* On World #AIDS Day (Dec. 1), help end mother-to-child #HIV transmission. Ask Congress to keep promise ... http://tr.im/wad_promise (VIDEO)

*Join another site or vote for organization.* The penultimate category involves asking followers to join another social media site or vote for the organization on another site. We found many organizations that use Twitter are also using other social networking tools. As in the following example, there is frequent cross-promotion across online social networking sites:

> *CatholicRelief:* Help nominate @CatholicRelief for "Best Nonprofit Use of Social Media" in 2009 #openwebawards (voting ends 11/15) &pls; RT! http://ow.ly/BIZs

*Learn how to help.* The final category is "Learn How to Help." This is different from directly asking for a donation because it sets up a two-step process: 1) learn how to help, 2) help. This is an important way to cultivate donors and other supporters. This category included indirect requests for a donation or other form of support. For example:

> *AmDiabetesAssn:* Happy Wednesday! Not feeling like you've been active enough this week? Take an action to #StopDiabetes!  http://bit.ly/36HcMS #diabetes

In sum, we found evidence of a variety of ways organizations are using Twitter to promote their organizations and mobilize supporters—what we refer to as "action" messages. Most commonly, organizations are directly asking for money or marketing specific events. Many organizations are also using Twitter to intermittently post job ads and calls for volunteers, as well as issue general "Learn how you can help" messages. Less commonly, organizations are engaging in direct product sales or attempting to mobilize supporters for lobbying and advocacy

HOW NONPROFIT ORGANIZATIONS USE SOCIAL MEDIA  17campaigns. Interestingly, nonprofits are also employing the promotion and mobilization function to help bolster the organizations' general "social media presence" on other platforms.

**Categorizing Organizations: Information Sources, Community-Builders, and Promoters**

In the previous section we classified the tweets made by our sample of organizations. Now we turn to an organizational-level analysis. Here, we attempt to build on these findings and make broader classifications of the organizations themselves based on their tweet profiles. Our goal here is to analyze each organization's aggregate tweet behavior to see if we can make some generalizations regarding different "types" of organizations. By examining the relative frequencies with which the organizations rely on the primary tweet functions, we can determine whether there exist, for instance, any truly "dialogic" organizations on Twitter.

To accomplish this, we examine organizations' relative reliance on the three main tweet functions—Information, Community, and Action. In order that each organization had sufficient tweets to enable an accurate assessment, for this analysis we used only organizations which we considered "active," measured as sending at least 3 tweets per week. If an organization is tweeting less often, their messages may get buried in their follower's feeds. Using this criterion, we found that 59 of the 73 organizations in our sample could be considered active.

Organizational variation in the emphasis of these three categories can be shown visibly in a ternary plot. Figure 1 is a ternary plot showing 59 dots, one for each of the "active" organizations, and is based on the proportion of each organization's tweets that are informational, community-building, and action-oriented, respectively (see the Web Appendix for a complete list with the relevant percentages for each organization). In a ternary plot, each organization's placement is determined by the relative proportions of each of the three types of tweets. A dot at the top vertex would indicate 100% of the organization's tweets are informational, a dot at the bottom left-hand vertex would indicate 100% action (promotional and mobilizational) tweets, and a dot in the bottom right-hand vertex would indicate 100% community-building tweets. Dots near the centroid of the triangle have roughly equal proportions of informational, community, and action-oriented tweets.

[Insert Figure 1 here]

The ternary plot shows us how heavily an organization relies on each of the primary functions. Based on this information, we denote three paradigmatic "types" of organizational users of Twitter: 1) "Information Sources," 2) "Community Builders," and 3) "Promoters & Mobilizers." The three dotted lines that meet at the centroid, carving the triangle into three equal sections, represent the approximate dividing lines separating the three types of organizations.

We see that only eight of the organizations are primarily "Community Builders," inasmuch as the plurality of their tweets are dialogic/community-building in nature, and that only four are chiefly "Promoters & Mobilizers." The bulk of the organizations (47) are primarily "Information Sources." There are three organizations (located directly on the left edge of the triangle) that have a mix of tweets that are informational and action-oriented in nature but none that are community-building; similarly, there are six organizations (located on the right edge) whose tweets are a mix of information and community but not action. There are none, however, located on the lower edge of the triangle, indicating that none of the organizations send out only a mix of community-building and action-oriented tweets.

This corroborates prior research that most organizations are not using social networking sites to their full dialogic, community-building potential (Greenberg & MacAulay, 2009; Waters, 2009). At the same time, the figure shows that almost all organizations *are* using dialogue as well as action, it is just rarely the predominant communicative purpose. Nevertheless, this is in contrast to what prior research has found with regard to organizational uses of websites. In short, though few organizations are fully "dialogic" in their use of social media, they are at least incorporating dialogue into their social media messages.

We should also be careful to say that organizations categorized as either "Community Builders" or "Promoters & Mobilizers" should not automatically be considered better than those relying on informational messages; instead of moving *beyond* the information function they may be *forgetting* the benefits that can accrue from using Twitter to send pertinent information to their followers. It is therefore possible that the organizations falling closer to the middle of the ternary plot are employing the most effective strategy, mixing information, dialogue, and promotion in equal parts. However, even this is likely an oversimplification. The more appropriate strategy may instead be the one that reflects the mission of the organization.

## Discussion and Conclusions

The advent of sophisticated, readily available social media applications, such as Facebook, del.icio.us, YouTube, and Twitter, has created hope that nonprofits will finally be able to fulfill the promise that the Web first created to engender a more active, "dialogic," and interactive civil society. To see whether organizations are tapping into this potential, we have examined how large US nonprofit organizations are using Twitter to engage with the public and their core stakeholders. Our study set out to accomplish two main tasks. First, based on an analysis of the 2,437 tweets sent by the organizations in our sample over a two-week period, we developed an original typology of organizational micro-blogging functions. Second, we then used this information to analyze and classify the organizations themselves based on their comparative reliance on the different micro-blogging functions.

For our first task, a tweet-level analysis, we began by classifying messages into 12

different categories. Based on our inductive analysis of the data, we then grouped these categories into three major functions; in our typology, each tweet serves primarily to either spread information, foster dialogue and build community, or mobilize supporters. We thus propose an "Information-Community-Action" micro-blogging message classification scheme.

For our second primary task, an organizational-level analysis, we classified organizations based on their tweet function utilization patterns. Our rough typology comprised three organizational types: "Information Sources," "Community Builders," and "Promoters & Mobilizers." We found that there are relatively few organizations in the latter two categories—in most organizations, the informational uses of Twitter predominate.

This represents the first study to analyze the content of nonprofit organizations' micro-blogging updates. More substantially, it is the first to classify social media messages by *organizations*, whether governmental, for-profit, or nonprofit; prior research has yet to attempt to classify organizational uses of either tweets or Facebook status updates. The nascent research in this area would benefit from such frameworks for understanding how organizations utilize new media in engaging with core stakeholders and the larger public. We thus help advance the literature in a critical area at the intersection of social media and organizational communication.

We found that, though dialogue is rarely the predominant form of communication, the overwhelming majority of organizations *are* using dialogue, community-building, and promotion and mobilization in their micro-blogging efforts. This finding leads us to propose that dialogue may *not* be the key form of social media-based organizational communication. Prior studies have implied that, in relying on informational communication, nonprofits have not been living up to their interactive, dialogic potential. The implication is that dialogue is the pinnacle of organizational communication. Instead, it may be that dialogue is simply one essential piece of the communication puzzle, and that information may always be the "base" form of communication. If this is correct, we would expect organizations, even fully "evolved" organizations, to continue to have more informational tweets than dialogic or action-oriented tweets. This is ultimately what our data show; whether it is a reflection of unfulfilled potential or the reality of a "hierarchy" of organizational communication functions is something to be tested.

One explanation for how the Information-Community-Action categories could represent a hierarchy of engagement is as follows. Sending information to stakeholders is important, and Twitter makes it easy to do this quickly and effectively. With only 140 characters, organizations have to be conscious of the main point they are trying to make. Users, in turn, can decide how much they want to know by either reading the tweet alone or, to learn more, by clicking on the links that are often included in organizational tweets. The second function, "community," involves dialogue and community building. This is where true engagement begins, when networks are developed and users can join in the conversation and provide feedback. The third category is "action," centering on marketing, promotion, and mobilization. This is where users do not just feel they are making a difference, but start doing something about it, whether it is showing up at an event, signing a petition, or making a donation. Organizations at this level are fully engaging their follower base. Users want information and to be part of the dialogue, but an organization fulfills its mission by getting its followers to do something for the cause it supports.

In this sense, the Information-Community-Action scheme might represent a "ladder" of organizational communication functions that would conform to the expectations of the resource mobilization perspective (McCarthy & Zald, 1977). Information could be seen as a core activity to attract followers, Community-focused messages serve to bind and engage a following of users, and Action-oriented messages serve to mobilize the resource—that is, the community—that has been developed through informational and community-oriented communication.

The above findings and ideas are potentially important, for they differ from prior research in what is the highest aim for social-networking-mediated organizational communication. Nonprofit researchers and stakeholder advocates may want, and have explicitly noted *dialogue* as the apex; however, for many (if not most) organizations, the apex may be promotion, marketing, and mobilization. This is not a normative assertion, but merely a recognition of what may be the implicit hierarchy in organizations' logic concerning the strategic potential of social media. Given the stark contrast with prior research, this is something that should certainly be subject to further study.

Our Information-Community-Action scheme is further important in how it goes beyond the simpler information/dialogue dichotomy that dominates prior research on new media and organizations. Our proposal of a distinct "action" function merits further attention. Roughly 16% of all messages in our data were promotional and mobilizational in nature. If these results are generalizable, future studies should endeavor to incorporate this dimension into their research. When seen together with the finding that a quarter of all messages were community-building in nature, we may need to update our understanding of organizational communication and stakeholder engagement in an age where social media continues to proliferate.

From a practical perspective, being on Twitter may in itself signal that an organization is willing to actively engage the public. However, such a signal will be more effective if reflected explicitly in the content of the actual messages sent. It is here that an organization needs to think about how Twitter can fit into its overall communication plan, as opposed to just thinking of tweeting as a trendy thing to do. The right way to use Twitter will not be found in following some formula, but in understanding an organization's needs and using the right tool to meet those needs. The first step is to understand what purpose each tweet serves as far as the public being targeted is concerned. The categories we have presented here help in this task inasmuch as they serve as a blueprint for understanding the core functions of micro-blogging messages.

We believe, moreover, that this categorization scheme has applicability beyond nonprofit organizations. Though for-profit firms' interest in informational and promotional and mobilizational messages is easily understood, it is not so obvious they would be interested in community-building and dialogue. Yet Rybalko and Seltzer (2010) found that 60% of *Fortune 500* firms on Twitter had dialogic features in their tweets, while Jansen et al. (2009) earlier found evidence of dialogue being a central feature of at least some companies' customer relationship management strategies. Jansen et al.'s (2009) additional case study of Starbucks provides evidence that Starbucks, for one, was highly interested in building an online *community*, and its Twitter feed in 2011 remains filled with community-building and dialogic messages. In brief, though it remains to be studied, we believe our "information-community-action" classification scheme will prove useful for studies of nonprofit, government, and for-profit organizations alike.

We also believe the categories are generalizable to other types of social media. For example, though Facebook has a larger range of functionality, Facebook statuses and tweets are so similar that many users, including several of the organizations in our study, send out the same messages on both outlets simultaneously. This will present challenges as well as opportunities for scholars studying the diffusion of information, viral marketing, and related phenomena, as well as those interested in the impact of messages on an organization's constituents.

Through these analyses, we have provided analytical insights into how organizations are using social media. This represents an important addition to the literatures on social media, on organizational communication, and on nonprofit organizations. Still, there is much that remains to be studied. To start, existing theory cannot fully account for the widespread organizational adoption of Twitter. Media richness theory (Daft & Lengel, 1986), which orders communication technologies according to their ability to facilitate shared meaning, is unable to explain why an organization would choose Twitter over Facebook, given that Facebook and other social media platforms offer a "richer" experience. In turn, "critical mass" theory (Markus, 1987), along with theories of social and institutional forces (Zorn, Flanagin, & Shoham, 2011), can explain why organizations *currently* feel pressure to adopt Twitter, now that it is close to standard practice, but not why micro-blogging in general rose to prominence, nor why Twitter ultimately dominated the field. In effect, theories regarding the unique *benefits* of Twitter for organizational communication have yet to be developed.

In this paper we have examined the messages organizations are sending on Twitter; future research should also look into how followers *respond* to these messages and otherwise try to participate in interactive conversations. For instance, many of the tweets asked followers to Retweet certain messages; future research could look into the implications of this practice. Similarly, many of the "community building" messages openly solicited responses from the organization's followers—future research could usefully examine if and how followers respond. Future research should also attempt to extend this paper by looking explicitly at the *interactions* between the organizations and their constituents—including the relationship between following and followers, the effects of tweet frequency on the growth in followers, and the effects on stakeholders of utilizing public reply messages, retweets, and private direct messages.

Though our sample was chosen to ensure that a large enough proportion of the organizations would have Twitter sites, one limitation of this study was the use of only the largest nonprofit organizations; future research might usefully employ a random sample of nonprofit organizations and/or look at a sample of smaller and mid-sized organizations. Given that so little research has looked at for-profit organizations' use of Twitter, and no articles published to date have looked at governmental uses of social media, there also exist great opportunities for studies that focus on these types of organizations as well.

Future research would also benefit from looking at which types of nonprofits rely more heavily on information, community-building, and action-oriented messages, respectively. We did not find important differences between the field in which the organization operated—such as arts, education, or health care—and the relative reliance on the different tweet functions. Consequently, future research may wish to investigate further the connection between specific aspects of the organizational mission and social media utilization. Overall, there is substantial room for more studies located at the intersection of organizations and social media.

We end on an exhortatory note. Although nonprofit organizations have become more interactive in their use of Twitter as opposed to their websites alone, we found Twitter is still used by many nonprofit organizations as an extension of information-heavy websites. These organizations are missing the bigger picture of its uses as a community-building and mobilizational tool. They are hence not using Twitter to its full capacity as a stakeholder-engagement vehicle. Being on Twitter is not enough—organizations need to know how to use the medium to fully engage stakeholders. Nevertheless, we found that an important minority of organizations *are* fully engaging their constituents through Twitter. More organizations need to follow their lead. Although it may seem counterintuitive that real interactions can happen in 140 characters or less, Twitter can in fact be used as a portal to substantive information, as a dialogic communication tool, and as a vehicle for the rapid mobilization of organizational followers.

HOW NONPROFIT ORGANIZATIONS USE SOCIAL MEDIA  26# References

Bortree, D., & Seltzer, T. (2009). Dialogic strategies and outcomes: An analysis of environmental advocacy groups' Facebook profiles. *Public Relations Review*, *35*, 317-19.

Briones, R. L., Kuch, B., Liu, B., & Jin, Y. (2011). Keeping up with the digital age: How the American Red Cross uses social media to build relationships. *Public Relations Review, 37*, 37-43.

Cameron, A.F., & Webster, J. (2005). Unintended consequences of emerging communication technologies: Instant Messaging in the workplace. *Computers in Human Behavior, 21,* 85-103.

Daft, R. L., & Lengel, R. H. (1986). Organizational information requirements, media richness, and structural determinants. *Management Science, 32*, 554-571.

D'Urso, S. C., & Pierce, K. M. (2009). Connected to the organization: A survey of communication technologies in the modern organizational landscape. *Communication Research Reports,* 26, 75-81.

Ellison, N., Steinfield, C., & Lampe, C. (2007). The benefits of Facebook "friends:" Social capital and college students' use of online social network sites. *Journal of Computer-Mediated Communication*, *12*, 1143-1168.

Greenberg, J., & MacAulay, M. (2009). NPO 2.0? Exploring the web presence of environmental nonprofit organizations in Canada. *Global Media Journal–Canadian Edition*, *2*, 63-88.

Hackler, D., & Saxton, G. D. (2007). The strategic use of information technology by nonprofit organizations: Increasing capacity and untapped potential. *Public Administration Review, 67*, 474-487.

Hughes, A.L., & Palen, L. (2009). Twitter adoption and use in mass convergence and emergency events. *International Journal of Emergency Management, 6,* 248-260.

HOW NONPROFIT ORGANIZATIONS USE SOCIAL MEDIA  27

HOW NONPROFIT ORGANIZATIONS USE SOCIAL MEDIA  28

**Table 1**  Tweet Functions

| Category | Example | Freq. | (%) |
|---|---|---|---|
| **Information** (58.6%) | | | |
| Information | *WorldVisionUSA:* Tribal clashes with police send 16,000 people fleeing from #Congo's Equateur Province ... http://tr.im/DRC_clashes #conflict | 1,429 | 58.6 |
| **Community** (25.8%) | | | |
| Giving recognition and thanks | *Smithsonian*: The @NationalZoo cuties are Twig Catfish. @UserID, @UserID & @UserID got it right!. More photos: http://ow.ly/Du3b | 321 | 13.2 |
| Acknowledgement of current & local events | *NYPL*: Patience and Fortitude Salute the Troops on Veterans Day! The Library Lions love a parade! http://bit.ly/2uvxm5 | 9 | 0.4 |
| Responses to reply messages | *DucksUnlimited*: @UserID We hope you get to go too. If you get out, tweet using the #duckhunting tag & let us know how it goes! | 199 | 8.2 |
| Response solicitation | *ChildFund*: Change a childhood #childfundcac event starts now. Give us your best tweets on child rights. Rules @ http://www.childfund.org/twitter | 99 | 4.1 |
| **Action** (15.6%) | | | |
| Promoting an event | *atAMNH*: Is there biology behind holiday madness? Find out what makes us naughty or nice at Dec. 2nd's SciCafe. RSVP on Facebook http://bit.ly/4bituI | 190 | 7.8 |
| Donation appeal | *UCPNational*: Sign up for "Black Friday" Deals on Amazon.com and a % of your purchase goes to UCP. Use this link: http://bit.ly/284BRx #disability #autism | 75 | 3.1 |
| Selling a product | *MetOpera*: The new online shop is now open! Browse through for great gifts, CDs, DVDs and more! http://www.metoperashop.org | 12 | 0.5 |
| Call for volunteers & employees | *ChildrensLA*: Pls RT Mission Critical:  Looking for a great online communications coordinator to help our hospital at http://bit.ly/9wRW4 #jobs | 20 | 0.8 |
| Lobbying and advocacy | *WorldVisionUSA*: On World #AIDS Day (Dec. 1), help end mother-to-child #HIV transmission. Ask Congress to keep promise ... http://tr.im/wad_promise (VIDEO) | 14 | 0.6 |
| Join another site or vote for organization | *CatholicRelief*: Were you at the Komen Global Race in DC this year? You definitely need to join the facebook group: http://bit.ly/qvSCk #globalrace (komenforthecure) | 29 | 1.2 |
| Learn how to help | *SalvationArmyUS*: Want to sign up for an Online Red Kettle, but need a little help? Here are some fun video tutorials to get you started! http://bit.ly/3j7GHN | 40 | 1.6 |
| | Total | 2,437 | 100% |

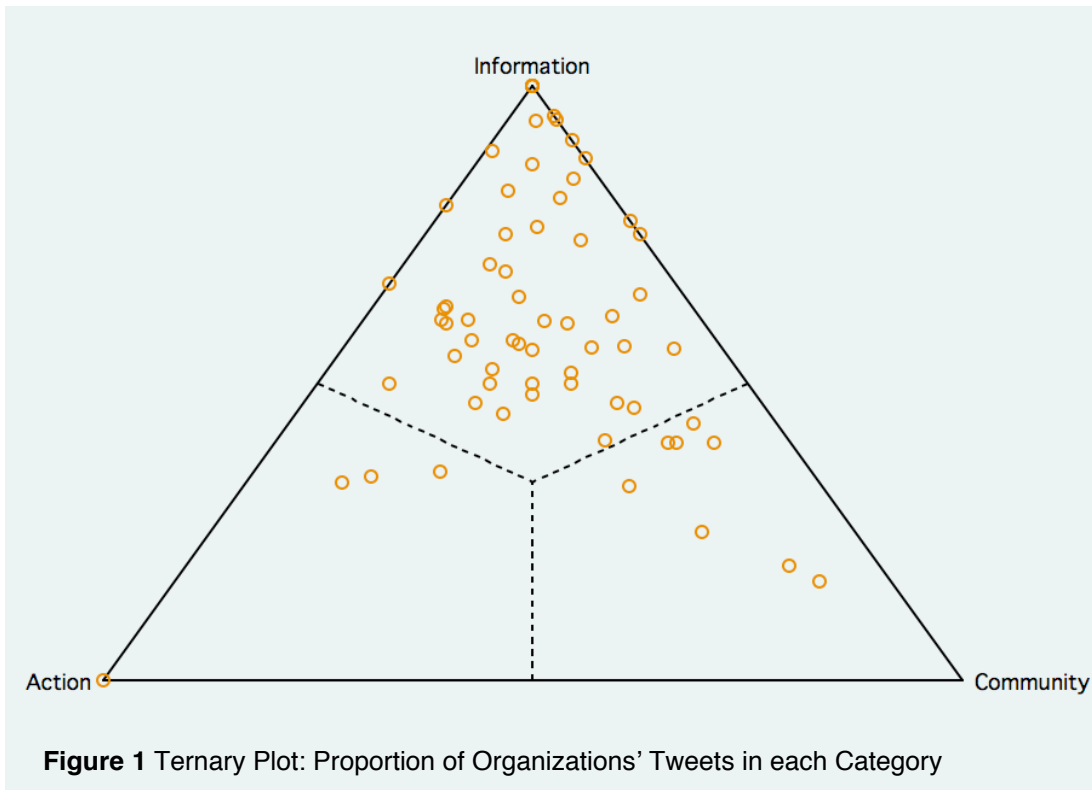

**Figure 1** Ternary Plot: Proportion of Organizations' Tweets in each Category

**About the Authors**

**Kristen Lovejoy** is a Ph.D. student at the University at Buffalo, SUNY. She completed her Masters in Communication and Leadership at Canisius College also in Buffalo, NY. She receives full funding through her teaching assistantship and is the President of the UB Communication Graduate Student Association. Her research interests include social media marketing, nonprofit organizations, and Galileo studies.

**Address:** University at Buffalo, North Campus, 359 Baldy Hall, Buffalo, New York 14260-1020. Email: kristenl@buffalo.edu.

**Gregory D. Saxton** is an Assistant Professor in the Department of Communication at the University at Buffalo, SUNY. His research interests are in organization-public relations, technology and management, and new media and organizational communication, concentrating on nonprofit organizations. His research has been published in *Public Relations Review, Social Science Quarterly, Public Administration Review,* the *British Journal of Political Science, Public Performance and Management Review*, *Nonprofit and Voluntary Sector Quarterly, International Interactions,* the *Journal of Peace Research,* and the *American Review of Public Administration*, among others.

**Address:** Department of Communication, 331 Baldy Hall, University at Buffalo, SUNY, Buffalo, NY 14260-1020. (Email) gdsaxton@buffalo.edu. (Tel) 716.645.1161. (Web) http://www.gregorysaxton.net.